\documentstyle [12pt] {article}

\catcode`@=12
\begin{document}
\thispagestyle{empty}
\begin{flushright} August 2000\
\end{flushright}
\mbox{}
\vspace{0.5in}
\begin{center}
{\Large	\bf Possible textures of neutrino and charged lepton mass matrices \\}
\vspace{1.0in}
{\bf Bipin R. Desai$^1$, Utpal Sarkar$^{1,2}$ and Alexander R. Vaucher$^1$\\}
\vspace{0.3in}
{$^1$ \sl Department of Physics, University of California, Riverside, 
California 92521, USA\\}
\vspace{0.1in}
{$^2$ \sl Physical Research Laboratory, Ahmedabad 380 009, India\\}
\vspace{1.0in}
\end{center}
\begin{abstract}\

We study a variety of different texture patterns in the basis in which
the charged lepton mass matrix or neutrino mass matrix or neither are
diagonal. The experimental results on the neutrinos provide sufficient
restrictions to allow only a small number of simple patterns. We discuss
particularly the texture zeroes which provide important relations among 
the mass eigenvalues and the mixing matrix. 

\end{abstract} 

\newpage
\baselineskip 18pt

\section{Introduction}

Recent results from superKamiokande on the atmospheric neutrino experiments
\cite{atm} has provided definite evidence for neutrino mass. It has
also been established that the solution of the atmospheric neutrino
anomaly requires a small mass squared difference between $\nu_\mu$
and $\nu_\tau$ with almost maximal mixing. Oscillations to sterile
neutrinos is ruled out at 90\% confidence level \cite{sobel}.
Another independent source of information on the mass of the 
neutrinos are the solar neutrino oscillations \cite{sol}. The 
latest results from
superKamiokande (mainly the day-night spectrum, which rules
out the small angle MSW solution and the vacuum oscillation 
solutions) \cite{suzuki} and the analysis of the solar neutrino data
\cite{soldata} have narrowed down the allowed region of the parameter
space to only one small region. Although the mixing angle is very 
large, maximal mixing is not allowed. The oscillation of $\nu_e$ 
into a sterile neutrino is also disfavoured.

\begin{table}[h]
\caption{Present experimental constraints on neutrino masses and mixing}
\begin{center}
\label{nuexp}
\begin{tabular}{||rcl||}
\hline \hline
&&\\
Solar Neutrino 
&:& $\Delta m^2 \sim (2.5 - 15) \times 10^{-5} eV^2$ \\
&& $ 0.25 < \sin^2 2 \theta < 0.65 ~~ \Longrightarrow ~~ 
0.26 < \sin \theta < 0.45 $ \\
&&\\
Atmospheric Neutrino 
&:&$ \Delta m_{\mu \tau}^2 \sim (1.5 - 5) \times 10^{-3} eV^2$ \\ 
&&$ \sin_{\mu \tau}^2 2 \theta > 0.88 ~~ \Longrightarrow ~~ 
\sin \theta_{\mu \tau} > 0.57  $\\
Neutrinoless  && \\
Double  Beta  Decay  &:&
$m_{\nu_e} < 0.2 eV$ \\
&&\\
CHOOZ  &:& $\Delta m_{e X}^2 < 7 \times 10^{-4} eV^2 $
for $\sin \theta_{eX} \sim 1$\\
&& or $\sin^2 2 \theta_{eX} < 0.1 ~~ \Longrightarrow ~~ 
\sin \theta_{eX} < 0.16 $ \\
&&\\
\hline \hline
\end{tabular}
\end{center}
\end{table}

In addition to these results, there are also some bounds from the
laboratory experiments. The long baseline reactor experiment, 
CHOOZ, has set stringent bounds on the disappeances of the $\nu_e$
\cite{chooz}.
In the situation under consideration with no sterile neutrinos, 
a small mass difference between $\nu_e$ and one of the combinations of
$\nu_\mu$ and $\nu_\tau$ can explain the solar
neutrino anomaly. The other combination will then have a mass squared
difference of the order of the $\nu_\mu$ and $\nu_\tau$ mass 
squared difference. For a mass squared difference of this amount,
the CHOOZ result gives a strong bound on the mixing angle. We also
assume that the neutrinos are Majorana particles, so that the 
lepton number violation at a large scale can also explain the
baryon asymmetry of the universe \cite{lepto}. This then implies
that there 
is constraint on the $\nu_e$ mass from the non-observation of the
neutrinoless double beta decay \cite{ndb}. All these constraints
are summarized in table \ref{nuexp}.

There have been several attempts to obtain textures of 
neutrino mass matrices and to relate them with the quark sector 
\cite{texture}, However, most of these studies have been based
on specific models that tried to get maximal mixing for the solar 
neutrino solution.

We will study the structures of the neutrino and charged lepton 
mass matrices in a model-independent way paying attention to any 
symmetry or regularity, particularly, the presence of any texture 
zeroes. We will also take into 
account the recent result that there is only one
solution to the solar neutrino problem, the one which does not allow
maximal mixing. As it turns out there are then very few choices 
that are left for us.

The atmospheric neutrino anomaly is explained by a maximal mixing of
$\nu_\mu$ and $\nu_\tau$. We thus assume that the mass eigenvalues 
$m_2$ and $m_3$ correspond to the states $\nu_2$ and $\nu_3$, which
are admixtures of the states $\nu_\mu$ and $\nu_\tau$. The mass 
squared difference required by the atmospheric neutrino anomaly
then can be written as, 
\begin{equation}
\Delta_{atm} = \Delta m_{\mu \tau}^2 = \Delta m_{23}^2 = m_3^2 - m_2^2 . 
\end{equation}
The corresponding mixing angle is written as, $\sin \theta_{\mu \tau}$,
which is almost maximal. The solar neutrino problem has a MSW solution
\cite{msw}, in which $\nu_1$ oscillates into $\nu_2$ with a mass
squared difference,
\begin{equation}
\Delta_{sol} = \Delta m_{e 2}^2 = \Delta m_{12}^2 = m_2^2 - m_1^2 . 
\end{equation}
The corresponding mixing angle is $s \sim \sin \theta_{e2}$. The third mass
squared difference is then determined in terms of these two mass
differences. The corresponding mixing angle $U_{e3}$ is then 
constrained by the CHOOZ data. 

We shall consider the mixing angle for the atmospheric neutrino 
anomaly to be maximal, $\sin \theta_{\mu \tau} \sim 1/\sqrt{2}$. However,
we will also consider $\sin \theta_{\mu \tau} < 0.57$ which is experimentally allowed.  
Similarly, if we assume, for simplicity, $U_{e3} = 0$, the neutrino mixing matrix becomes,
\begin{equation}
U_\nu = \pmatrix{ c& -s & 0 \cr {s \over \sqrt{2}} & 
{c \over \sqrt{2}} & - {1 \over \sqrt{2}} \cr
{s \over \sqrt{2}} & {c \over \sqrt{2}} & {1 \over \sqrt{2}}}
\label{uai}
\end{equation}
where $c = \sqrt{1 - s^2}$. With this restrictive
mixing matrix we find that there is no neutrino mass matrix with any texture
zeroes (in the basis where the charged lepton is diagonal). For exact bi-maximal solutions, {\it i.e.}, with
$s = c = 1/\sqrt{2}$ there are solutions with texture zeroes,
but when we consider the present bound on $s$, there are no
solutions. So, to generalise our analysis we would also allow 
$U_{e3}$ to be non-vanishing, but constrained to be small. 

In our present analysis we shall not include CP violations. The
mass matrices and the mixing matrices would then be real.
For mass matrices, $M_e$ and $M_\nu$ and the 
corresponding charged current interaction, the Lagrangian is given by,
\begin{equation}
{\cal L} = \frac{g}{\sqrt{2}} \overline{\ell}_{m L} ~
\gamma^\mu ~\nu_{m L} ~
W_\mu^- - \overline{\ell}_{m L}~ M^\prime_{e m n} ~\ell_{n R} - 
{\nu}_{m L}~ M_{\nu mn} ~\nu_{nL} .
\end{equation}
The charged lepton mass matrix is not symmetric, but can be 
diagonalised by a bi-unitary transformation
$$V^\dagger_{e \alpha m R}
M^\prime_{e mn} V_{e n \beta L} = M_{e \alpha} \delta_{\alpha \beta },$$
where $\alpha, \beta = e, \mu, \tau $ are the physical states. 
Since the right handed mixing matrix $V_{e m \alpha R}$ does not
enter into the charged current interactions, we can write
$V_{em\alpha R} = K_{e m n} V_{m \alpha L}$ and without loss of
generality symmetrize the charged lepton mass matrix. The new
symmetric charged lepton mass matrix, $M_{e m n} = M^\prime_{e m l}
K_{e l n}$ can now be diagonalised by the unitary matrix,
$$V^\dagger_{e \alpha m L}
M_{e mn} V_{e n \beta L} = M_{e \alpha} \delta_{\alpha \beta }.$$
The Majorana mass matrix of the neutrinos is always symmetric and
hence can be diagonalised by a unitary transformation,
\begin{equation}
V^\dagger_{\nu i m} M_{\nu mn} V_{\nu n j} = 
M_{\nu i} \delta_{i j} .
\end{equation}
Since we are working with real matrices, all unitary matrices can
be replaced by orthogonal matrices and we have not introduced the
Majorana phase matrix. 

In this flavour basis, when the charged lepton mass matrix is
diagonal, the charged current interaction contains the mixing
matrix $U_{\nu \alpha i}$ which is given in equation (\ref{uai}). 
This mixing matrix $U_{\nu \alpha i}$, which is measured from
experiment can be written in terms of the unitary matrices which
diagonalise the mass matrices $V_{e m \alpha}$ and $V_{\nu i m}$
as,
\begin{equation}
U_{\nu \alpha i} = V_{e \alpha m}^T V_{\nu m i}  .
\end{equation}
This enters in the charged current interactions of the
physical charged lepton states with the neutrino mass eigenstates.
This is analogous to the Cabibbo-Kobayashi-Maskawa 
matrix of the quark sector. 
In the basis in which the charged lepton mass matrix is diagonal,
the mixing matrix is simply $U_\nu = V_\nu$, while in
the basis in which the neutrino mass matrix is diagonal, the mixing
matrix is given by, $U_\nu = V_e^T$.

\begin{table}
\caption{Neutrino mass matrices and the corresponding mixing matrices
for some representative choice of parameters with $M_e$ diagonal. 
In all these cases, the 
mass differences lie inthe range $\Delta m_{12}^2 \sim (4.5-8) \times 
10^{-5}$ eV$^2$ and $\Delta m_{23}^2 \sim (3.2-3.6) \times 
10^{-3}$ eV$^2$.}
\begin{center}
\label{numix}
\begin{tabular}{||lcllc||}
\hline \hline
&Mass matrices& Parameters &Masses in eV& Mixing matrices\\ 
\hline
&&&&\\
A& $\pmatrix{0 & a & a \cr a & 0 & b \cr a & b 
& 0} $ &
$ \begin{array}{l} a=0.04 \\ b=0.00025 \end{array}$ &
$ \begin{array}{l} m_1 = 0.05682 \\ m_2 = -0.05632 \\ m_3 = -0.00025 
\end{array} $&
$\pmatrix{0.71&0.71& 0 \cr 0.50& 0.50 & 0.71
\cr 0.50 & 0.50 & 0.71}$ \\
&&&&\\
B& 
$\pmatrix{a & b & b \cr b & b & -b \cr b & -b & b}$ & 
$ \begin{array}{l} a=0.00025 \\ b=0.04 \end{array}$ &
$ \begin{array}{l} m_1 = 0.05682 \\ m_2 = -0.05632 \\ m_3 = 0.08 
\end{array} $&
$\pmatrix{0.71&0.71& 0 \cr 0.50& 0.50 & 0.71
\cr 0.50 & 0.50 & 0.71}$ \\
&&&&\\
C& 
$\pmatrix{a & a & a \cr a & b & -b \cr a & -b & b}$ & 
$ \begin{array}{l} a=0.005 \\ b=0.03 \end{array}$ &
$ \begin{array}{l} m_1 = 0.01 \\ m_2 = -0.005 \\ m_3 = 0.06 
\end{array} $&
$\pmatrix{0.82&0.58& 0 \cr 0.41& 0.58 & 0.71
\cr 0.41 & 0.58 & 0.71}$ \\ 
&&&&\\
D& 
$\pmatrix{0 & a & a \cr a & b & -b \cr a & -b & c}$ & 
$ \begin{array}{l} a=0.0014 \\ b=0.025 \\ c=0.04 \end{array}$ &
$ \begin{array}{l} m_1 = -0.0011 \\ m_2 = 0.007 \\ m_3 = 0.057 
\end{array} $&
$\pmatrix{0.93 & 0.36 & 0.007 \cr 0.28 & 0.74 & -0.60
\cr 0.22 & 0.56 & 0.80}$ \\ 
&&&&\\
E& 
$\pmatrix{0 & a & 0 \cr a & b & b \cr 0 & b & c}$ & 
$ \begin{array}{l} a=0.003 \\ b=0.03 \\ c=0.04 \end{array}$ &
$ \begin{array}{l} m_1 = 0.0009 \\ m_2 = 0.0055 \\ m_3 = 0.065 
\end{array} $&
$\pmatrix{0.92 &0.38& 0.03 \cr 0.30 & 0.70 & 0.65
\cr -0.22 & -0.61 & 0.76}$ \\
&&&&\\
F& 
$m \pmatrix{2 s^4 & \sqrt{2} s^3 & \sqrt{2} s^3 \cr 
\sqrt{2} s^3 & 1 + s^2 & s^2 - 1 \cr \sqrt{2} s^3 & s^2 -1 & 1 + s^2}$ & 
$ \begin{array}{l}  m = 0.03 \\
s = -0.375  \end{array}$ &
$ \begin{array}{l} m_1 = 0 \\ m_2 = 0.009 \\ m_3 = 0.06 
\end{array} $&
$\pmatrix{-0.93 &0.35 & 0 \cr 0.25 & 0.66 & -0.71
\cr 0.25 & 0.66 & 0.71}$ \\ &&&&\\
\hline \hline
\end{tabular}
\end{center}
\end{table}

\section{Diagonal Charged Lepton Mass Matrix}

We shall first consider the neutrino mass matrices in the basis
in which the charged lepton mass matrix is diagonal. A general form
of the mass matrix, which gives the mixing matrix of equation (\ref{uai}),
may be obtained from a diagonal neutrino mass, $M_\nu^{diag} = 
{\rm Diag.} \{m_1, m_2, m_3\} $ as,
\begin{equation}
M_e =M_e^{diag} 
\end{equation}
and
\begin{equation}
M_\nu  = U_{\nu \alpha i} M_{\nu i i}^{diag} U_{\nu i \beta}^T
= \pmatrix{  m + \Delta \cos 2 \theta 
&  {1 \over \sqrt{2}} \Delta \sin 2 \theta 
&  {1 \over \sqrt{2}} \Delta \sin 2 \theta  \cr 
{1 \over \sqrt{2}} \Delta \sin 2 \theta & 
{1 \over 2} \left( M - \Delta \cos 2 \theta  \right) & 
{1 \over 2} \left( \Delta_3 - \Delta \cos 2 \theta  \right) \cr 
{1 \over \sqrt{2}} \Delta \sin 2 \theta & 
{1 \over 2} \left( \Delta_3 - \Delta \cos 2 \theta  \right) & 
{1 \over 2} \left( M - \Delta \cos 2 \theta  \right) } ,
\end{equation}
where, $m = (m_1 + m_2)/2$, $\Delta = (m_1 - m_2)/2$, $M =
m + m_3$, $\Delta_3 = m - m_3$ and $\cos 2 \theta = c^2 - s^2$.

Although this matrix does not allow any texture zeroes, it is 
instructive in the sense that it gives $M_{12} = M_{13}$ and 
$M_{22} = M_{33}$. Moreover, if we require an inverted hierarchy
with a very small $m_{3}$ then we also have $C=D$. Using such
symmetries, several forms of the neutrino mass matrix can be
obtained. 
We list a few possible mass matrices in table 2 along with
the mixing matrix and eigenvalues for a representative set of parameters.

The mass matrix 
$$ \pmatrix{0 & a & a \cr a & 0 & b \cr a & b & 0 } $$
has been studied \cite{ag}, that has texture zeroes along the 
diagonal. This is pattern (A) in table 2. For the choice, $m_1 \approx 
-m_2$, we have $\Delta \neq 0$ and $m \approx 0$ 
and, therefore, for the (11) element to vanish we must
have $\cos 2 \theta \approx 0$. 
This leads to bi-maximal
mixing matrix. Although this has maximum number of texture zeroes,
this mass matrix cannot give non-maximal mixing angle indicated by the
solar neutrino problem. 

For another simple pattern given in (B) in table 2,
$$ \pmatrix{a & b & b \cr b & b & -b \cr b & -b & b } ,$$
the choice of $m_1 \approx -m_2$ (therefore $\Delta \neq 0$)
and $m_3 \gg m_1$ leads to $m \approx 0,~ M \approx m_3$, and
$\Delta \approx -m_3$. Therefore, for (22) and (23) matrix elements
to have opposite signs as above, we once again have $\cos 2
\theta \approx 0$, which leads to bi-maximal mixing. A
judicious choice of the parameters $m_1, ~m_2$ and $m_3$ can
then be made to have the (12) and (22) elements the same.

Pattern (C), because of the choice of $m_1$, $m_2$, $m_3$, is an
example which does not allow bi-maximal mixing. The simplest of
the mass matrices with one and two texture zeroes are given,
respectively, by
$$ \pmatrix{0 & a & a \cr a & b & -b \cr a & -b & c } $$
which is pattern (D), and 
$$ \pmatrix{0 & a & 0 \cr a & b & b \cr 0 & b & c } $$
which is pattern (E). We note that in order to to fit the neutrino data 
(particularly, $\theta_{\mu \tau} \approx 45^\circ$), the structure of 
both the matrices must be such that $c \neq b$. That is, the
(33) element is no longer equal to the (22) element as implied by
(8). As a consequence, one finds that neither is $U_{e 3} = 0$. If 
$U_{e 3}$ were to vanish then there is no solution with (non-diagonal) 
texture zeroes.

\section{Diagonal neutrino mass matrix}

We next consider the case where the charged lepton mass matrix
is not diagonal. The simplest situation would be where the 
neutrino mass matrix is diagonal and the entire mixing matrix comes from 
the diagonalisation of the charged lepton mass matrix. This is what
happens in the case of democratic mass matrix \cite{frx}, in which 
case the charged lepton mass matrix is democratic with rank one. 
However, in that case the mixing angle for the solar neutrino case
comes out to be maximal, which is now ruled out at the 95\% confidence
level. Since the neutrino mass matrix is diagonal, the mixing 
matrix is the transpose of the matrix diagonalising the charged lepton
mass matrix, 
\begin{equation}
V_e = U^T_{\nu} = \pmatrix{  c&  {s \over \sqrt{2}} & 
{s \over \sqrt{2}} \cr -s & {c \over \sqrt{2}} &
{c \over \sqrt{2}} \cr 0 &  - {1 \over \sqrt{2}} & {1 \over \sqrt{2}}} .
\label{ubi}
\end{equation} 
Using this mixing matrix, we can now write down the most general
charged lepton mass matrix (pattern 1 in table 3) in terms of the 
mass eigenvalues as,
\begin{equation}
M_\nu = M_\nu^{diag} ~~~~~ {\rm and} ~~~~~
M_e = V_e M_e^{diag} V_e^T = \pmatrix{ m_+ s^2 + m_e c^2 &
(m_+ - m_e) sc & m_- s \cr (m_+ - m_e) sc & m_+ c^2 + m_e s^2 &
m_- c \cr m_- s & m_- c & m_+ }
\end{equation}
where $M_e^{diag} = {\rm Diag.} \{ m_e, m_\mu, m_\tau \}~$,
$~m_+ = (m_\mu + m_\tau)/2$ and $m_- = (m_\mu - m_\tau)/2$. 

\begin{table}
\caption{Patterns for $M_\nu$ and $M_e$ for different textures which
are consistent with the experimental values of the mixing 
matrix $U_\nu$. We defined $m_\pm = (m_\mu \pm m_\tau)/2$, $m_{e \tau}
= \sqrt{m_e m_\tau}$, $m_{e \mu} = \sqrt{m_e m_\mu}$, $\tilde m =
\sqrt{2} \Delta s_{12}$ and $\tilde m_3 = {\Delta_3 \over 2} s_{12}$
and considered $s \approx 0.35$.}
\begin{center}
\label{e-mix}
\begin{tabular}{||lccc||}
\hline \hline
& $M_\nu$ &
 $M_e$ & $U_\nu$ \\
\hline
&&&\\
1$^\ast$& $\pmatrix{m_1 & 0 & 0 \cr 0 & m_2& 0 \cr 0 & 0 & m_3}$&
 $\pmatrix{ a & b & m_- s \cr b & d &
m_- c \cr m_- s & m_- c & m_+  }$& same as (3) \\
&&&\\
2&$ {m_2 \over 2} \pmatrix{s^2 & sc & 0 \cr
sc & c^2 & 0 \cr 0 & 0 & {2 m_3 \over m_2}} $& $ 
\pmatrix{m_e & 0 & 0 \cr 0 & m_+ & m_- \cr 0 & m_- & m_+} $ & same as (3) \\
&&&\\
3$^\dag$& ${m_2 \over 2} \pmatrix{s^2 & sc & 0 \cr
sc & c^2 & 0 \cr 0 & 0 & {2 m_3 \over m_2}} $& $ m_0
\pmatrix{ 0 & a & a \cr a & b & b \cr a & b & d} $ &
$  \pmatrix{ -0.93 & 0.37 & 0 \cr -0.28 & -0.70 & 
0.66 \cr 0.24 & 0.61 & 0.75} $ \\
&&&\\
4$^\ddag$&${m_2 \over 2} \pmatrix{s^2 & sc & 0 \cr
sc & c^2 & 0 \cr 0 & 0 & {2 m_3 \over m_2}} $& $ m_0
\pmatrix{ 0 & a & 0 \cr a & b & b \cr 0 & b & d} $ &
$ \pmatrix{ -0.95 & 0.30 & 0.05 \cr 0.19 & 0.72 & -0.67 \cr 
-0.24 & -0.63 & -0.74} $ \\
&&&\\
5& $\pmatrix{ m - \tilde m & {\Delta \over \sqrt{2}} 
+ \tilde m_3 & {\Delta \over \sqrt{2}} - 
\tilde m_3 \cr {\Delta \over \sqrt{2}} + 
\tilde m_3 &  {M \over 2} + \tilde m &
 {\Delta_3 \over 2} +  {\tilde m \over {2}} \cr
{\Delta \over \sqrt{2}} - \tilde m_3 &
{\Delta_3 \over 2} +  {\tilde m \over {2}} &
{M \over 2} }$ \hskip -.1in & 
$\pmatrix{ 0 & m_{e \mu} & 0 \cr m_{e \mu} & m_\mu & 
m_{e \tau} \cr 0 & m_{e \tau} & m_\tau}$
& same as (3) \\
&&&\\
6& $\pmatrix{ m + {\Delta^2 \over \Delta_3}& 0 & 
{3 \Delta \over 2 \sqrt{2} } \cr 0 & {M \over 2} - {\Delta^2 \over \Delta_3} & 
{\Delta_3 \over 2} - {\Delta^2 \over 2 \Delta_3} \cr 
{3 \Delta \over 2 \sqrt{2} }  \hskip -.1in &
{\Delta_3 \over 2} - {\Delta^2 \over 2 \Delta_3} & {M \over 2} }$&
$\pmatrix{ 0 & m_{e \mu} & 0 \cr m_{e \mu} & m_\mu & 
m_{e \tau} \cr 0 & m_{e \tau} & m_\tau}$
&  same as (3) \\
&&&\\
\hline \hline
\multicolumn{3}{l}{$^\ast$ $a = m_+ s^2 + m_e c^2,~b=(m_+ - m_e) s c$, and
$d= m_+ c^2 + m_e s^2$.} \\
\multicolumn{3}{l}{$^\dag$ $m_0 = 0.82 ~{\rm GeV},~a=0.025,~b=1$ and $d=1.285.$ }\\
\multicolumn{3}{l}{$^\ddag$ $m_0 = 1 ~{\rm GeV},~a=0.01,~b=0.9$ and $d=1.1.$  }\\

\end{tabular}
\end{center}
\end{table}

{}From the above expressions a few points become clear. None of the elements 
in $M_e$ 
could vanish and be consistent with the above mixing matrix. The smallness of
the electron mass $m_e$ implies that the various elements of the
mass matrix have to be given very precisely in order to get, simultaneously, 
the mass hierarchy between the charged lepton masses and the required
mixing matrix. So, this form of the mass matrix is
unlikely to predict any texture zeroes. In other words, it is not possible 
for texture zeroes to exist without invoking new 
parameters, which have to be fine tuned to get the required mixing 
matrix. A simple vanishing of any elements of
the mass matrix will, therefore, not allow us to get the required mixing matrix.

\section{Genaral textures}

Let us consider the case where one of the mixing angles
comes from the charged lepton mass matrix and the other from the 
diagonalisation of the neutrino mass matrix \cite{yan}. If we now require 
that the mixing angle of the atmospheric neutrino anomaly comes 
from the neutrino sector and the (almost) maximal mixing angle for 
the solar neutrino comes from the charged lepton mixing matrix, then
again we need fine tuning. For any of the elements of the charged
lepton mass matrix to vanish, it is not possible to get a 
solution, since the maximal mixing in the neutrino sector would then make
the $U_{e3}$ element large. The only possibility, therefore, is to get 
maximal mixing between $\nu_\mu$ and $\nu_\tau$ from the charged 
lepton sector. In this case, in the above charged lepton mass matrix
we can put $s=0$ allowing the mixing in the $\nu_e$ to $\nu_\mu$ 
sector to come from the neutrino sector. The mass matrices in this
case are given by,
\begin{equation}
M_e = \pmatrix{m_e & 0 & 0 \cr 0 & m_+ & m_- \cr 0 & m_- & m_+}~;
~~~~ {\rm and} ~~~~M_\nu = {m_2 \over 2} \pmatrix{ s^2 & sc & 0 \cr
sc & c^2 & 0 \cr 0 & 0 & {2 m_3 \over m_2}} .
\end{equation}
This is pattern 2 in table 3.
The corresponding unitary matrices, which diagonalise these matrices
are,
\begin{equation}
V_e = \pmatrix{1 & 0 & 0 \cr 0 & {1 \over \sqrt{2}} & {1 \over \sqrt{2}}
\cr 0 &  -{1 \over \sqrt{2}} &  {1 \over \sqrt{2}}} ~~~~ {\rm and} ~~~~
V_\nu = \pmatrix{c & -s & 0 \cr s & c & 0 \cr 0 & 0 & 1}
\end{equation}
which gives the required neutrino mixing matrix of equation (\ref{uai}) 
$U_\nu = V_e^T V_\nu $. 

We shall now check which of the charged lepton mass matrices resemble
the textures observed in the quark sector where two prominent patterns 
have been observed \cite{mmfx,desai}
\begin{equation}
\pmatrix{0 & * & * \cr * & * & * \cr * & *& *} ~~~~~ {\rm or}
~~~~~ \pmatrix{0 & * & 0 \cr * & * & * \cr 0 & *& *} .
\end{equation}
We shall not impose any condition on the neutrino mass matrices, 
since the origin of the neutrino mass matrix appears quite different from
the up quark sector. So, we assume the same neutrino mass matrix
as above and obtain the form of the charged lepton mass matrix which
can give us the required mixing matrix and the mass eigenvalues.
There exists two possible charged lepton
mass matrices corresponding to these two textures, which are,
\begin{equation}
M_e^A = m_o \pmatrix{ 0 & a & a \cr a & b & b \cr a & b & c} ~~~~~ {\rm or}
~~~~~ M_e^B = m_o \pmatrix{ 0 & a & 0 \cr a & b & b \cr 0 & b & c} .
\end{equation}.

We point out that the (33) elements above do no follow the pattern given in (8)
which, therefore, allows the texture zeroes to develop in (14). These
two patterns (patterns 3 and 4) are given in table 3 for some typical  
parameters.

\section{Comparison with the Quark Texture}

We will now take a closer look at the quark sector by comparing the
individual matrix elements. From the analysis of the textures of the 
$u-$ and $d-$quark mass matrices it is known that an excellent 
representation for the $d-$quark is given by \cite{mmfx,desai},
\begin{equation}
D= \pmatrix{0 & \sqrt{m_1 m_2} & 0 \cr \sqrt{m_1 m_2} & m_2 & 
\sqrt{m_1 m_3} & \cr 0 & \sqrt{m_1 m_3} & m_3}
\end{equation}
where the masses are in hierarchical order with $m_1=m_d$,
$m_2=m_s$ and $m_3=m_b$. This matrix correctly reproduces $V_{us}$ and
$V_{cb}$ of the CKM matrix as,
\begin{equation}
V_{us} \approx \sqrt{m_d \over m_s} , ~~~~~~~~~ V_{cb} 
\approx \sqrt{m_d \over m_b}
\end{equation}
If we assume that the charged-lepton mass matrix has an identical form to 
$D$, then we can write an analogous relation 
\begin{equation}
M_e = \pmatrix{ 0 & \sqrt{m_e m_\mu} & 0 \cr \sqrt{m_e m_\mu} & m_\mu & 
\sqrt{m_e m_\tau} \cr 0 & \sqrt{m_e m_\tau} & m_\tau}
\end{equation}
Unlike the $D$, several different representations are possible for $U$. 
Furthermore, it is not at all clear, because of neutrino's Majorana
character and the possible presence of a see-saw mechanism, that a 
similarity exists between the neutrino and the $u-$sector. Therefore
we write, with (17) as the basis for $M_e$, the most general matrix 
for $M_\nu$
\begin{equation}
M_\nu = \pmatrix{a & b & c \cr b & d & e \cr c & e & f}
\end{equation}
We will obtain $M_\nu$ above by comparing it to (8) after we 
diagonalize $M_e$ in (17) since (8) is actually in the basis with $M_e$
diagonal. 

We write 
\begin{equation}
M_e^{diag} = T^\dagger M_e T
\end{equation}
where $M_e$ is given by (17) and $T$ is given, in the small angle 
approximation as
\begin{equation}
T = \pmatrix{ 1 & s_{12} & 0 \cr - s_{12} & 1 & s_{23} \cr
0 & -s_{23} & 1}
\end{equation}
where
\begin{equation}
s_{12} \approx \sqrt{ m_e \over m_\mu} , ~~~~~~~~ 
s_{23} \approx \sqrt{ m_e \over m_\tau}
\end{equation}
Thus in the representation where $M_e$ is diagonal 
\begin{equation}
M_e = M_e^{diag}
\end{equation}
one can express $M_\nu$ as 
\begin{equation}
M_\nu = T \pmatrix{a & b & c \cr b & d & e \cr c & e & f} T^\dagger
\end{equation}
and obtain the elements $a,b,..$ etc, by comparing the above 
expression with (8).
Since $s_{23} \ll s_{12}$ from (21), we will keep only $s_{12}$ to 
obtain (with $\theta = 45^\circ$, for simplicity)
\begin{eqnarray}
a&=& m - \sqrt{2} \Delta s_{12} \nonumber \\
b&=& {\Delta \over \sqrt{2}} + {\Delta_3 \over 2} s_{12}
\nonumber \\
c&=& {\Delta \over \sqrt{2}} - {\Delta_3 \over 2} s_{12} \nonumber \\
d&=& {M \over 2} + \sqrt{2} \Delta s_{12} \nonumber \\
e&=& {\Delta_3 \over 2} +  {\Delta \over \sqrt{2}} s_{12} \nonumber \\
f&=& {M \over 2} 
\end{eqnarray}
This is pattern 5 in table 3.
Once again there are no texture zeroes in $M_\nu$ unlike the $u-$sector
where, as we mentioned previously,
several representations with two or three texture zeroes are found
\cite{desai} for the same pattern in the $d$- and, therefore, in the
$M_\nu$ structures.

What if we demand a texture zero in $M_\nu$ ? The only possible spots 
that would physically make sense are the (12) (and (21)), or (13) (and (31))
locations in the mass matrix (23) in which case we find
\begin{equation}
\Delta = \pm  {\Delta_3 \over \sqrt{2}} s_{12} 
\end{equation}
From (8) and (21) one can express this relation as
\begin{equation}
\Delta_{12}^2 = \left( {2 \sqrt{2} m \over m + m_3} \right)
\sqrt{m_e \over m_\mu} \Delta_{23}^2
\end{equation}
where $m$ is the average mass of $m_1$ and $m_2$ and $\Delta_{12}^2$
and $\Delta_{23}^2$ are defined in (1) and (2). 

If we take a hierarchical
structure for the neutrino masses and take $m_1=0$ then from (1),
(2) and Table 1 we find that the above relation is, indeed, quite
well satisfied. If we choose the (12) (and (21)) element to 
vanish then the matrix is as given by pattern 6 in table 3.

These are the only simple forms of the charged lepton masss matrices allowed 
which can give the mixing angles for the solar neutrino problem, when
the atmospheric neutrino mixing comes from the diagonalisation of the 
neutrino mass matrix. This completes the possible forms of the mass
matrices which are consistent with the present experiments.

In summary, we have presented different possible forms of the neutrino and
charged lepton mass matrices which have simple patterns and
which can give us the required mixing
matrix and the mass eigenvalues. There are very few texture zero 
forms of the mass matrices which are still consistent. In all the cases
some of the elements are required to be equal to each other. If we
demand that the charged lepton mass matrix be same as the 
down quark mass matrix in texture, then some interesting results emerge, 
particularly if we also demand texture zeroes in the neutrino mass
matrix. In the basis in which the charged lepton mass matrix is
diagonal there are very few solutions with textures zeroes. 

~\vskip 0.5in
\begin{center} {ACKNOWLEDGEMENT}
\end{center}

One of us (US) would like to thank Ernest Ma for his hospitality 
at the University of California at Riverside. 
We acknowledge discussions with Ernest Ma and 
G. Rajasekaran. This work was supported in part by the
U.S. Department of Energy under contract No. DE-FG03-94ER40837.

\newpage
\bibliographystyle{unsrt}

\end{document}